\documentclass[pra,onecolumn,showpacs,showkeys,superscriptaddress]{revtex4}
\usepackage{comment}
\usepackage{graphics}
\usepackage{epsfig}
\newcommand{\be}{\begin{equation}}
\newcommand{\ee}{\end{equation}}
\newcommand{\ba}{\begin{eqnarray}}
\newcommand{\ea}{\end{eqnarray}}
\begin{document}
\title{Quantum mechanical description of Stern-Gerlach experiments}
\author{G. Potel}
\affiliation{Departamento de F\'\i sica At\'omica, Molecular y Nuclear, 
Aptdo. 1065, 41080 Sevilla, Spain}
\author{F. Barranco}
\affiliation{Departamento de F\'\i sica Aplicada 3, E.S.I. 
Isla de la Cartuja, Sevilla, Spain }
\author{S. Cruz-Barrios}
\affiliation{Departamento de F\'\i sica Aplicada 1, E.U.P. 
Virgen de África , Sevilla, Spain } 
\affiliation{Departamento de F\'\i sica At\'omica, Molecular y Nuclear, 
Aptdo. 1065, 41080 Sevilla, Spain}
\author{J. G\'omez-Camacho}
\affiliation{Departamento de F\'\i sica At\'omica, Molecular y Nuclear, 
Aptdo. 1065, 41080 Sevilla, Spain}
\begin{abstract}
The motion of neutral particles with magnetic moments in an inhomogeneous 
magnetic field is described in a quantum mechanical framework.
The validity of the semi-classical approximations which are generally used to 
describe these phenomena is discussed. Approximate expressions for the 
evolution operator are derived and compared to the exact calculations.
Focusing and spin-flip phenomena are predicted.
The reliability  of Stern-Gerlach 
experiments to measure spin projections is assessed
in this framework.
\end{abstract}
\maketitle
\section{Introduction}

The Stern-Gerlach experiment consists in taking a beam of particles
that have a neutral electric charge, but a finite magnetic moment, and
passing them through an inhomogeneous magnetic field. The observed
result is that the particles deflect differently depending on the spin
projection along the magnetic field. So, by measuring the deflection, one can
infer the value of the spin projection of the particles along the direction 
of the magnetic field.  The description of this phenomenon is done with the
following assumptions:

i) The spin projection along the z-axis, taken along the magnetic field at the
centre of the beam, is conserved.

ii) Particles with different spin projections along the z-axis, as they go 
through the 
inhomogeneous magnetic field, suffer a force in the 
z-direction that is given by the product of the magnetic moment times the 
gradient of the field times the spin projection.

This is what we will call the {\em text-book} description
of the Stern-Gerlach experiment \cite{eis,lev,mer,mes}.
Thus,  considering the particle 
position as a pointer, and the spin projection as the quantum property to be 
measured,  the Stern-Gerlach setup is associated to a measurement 
operator on the spin state which has as eigenvalues the spin projections
along the z-axis. Under the  {\em text-book} description, the Stern Gerlach 
experiment corresponds to an ``ideal'' measurement, in the sense of von 
Neumann 
\cite{neumann},
 because the quantum state is not modified by the measurement
process when it is an eigenstate of the measuring apparatus. 
Besides, it is ``completely reliable'', in the sense discussed in 
\cite{bassi}, because the the position is completely correlated with the spin
projection.

However, when the experiment is investigated in more detail,
the situation becomes more complicated. As the magnetic field has zero 
divergence,
then it is not possible to have a gradient of the field only in one 
direction. This produces terms in the Hamiltonian that can change the spin
of the incident particle. A detailed investigation of these effects was made
in a recent publication \cite{prasara}, making use of the concept of
coherent internal states \cite{sara} in a  semi-classical 
approach. The main results of that paper was that, indeed, when a beam of 
particles goes through a Stern-Gerlach magnet, the different spin projections
deviate depending on the spin projection. However, when the size of the beam
is not very small compared to the range of inhomogeneity of the magnetic field,
additional effects occur. 

i) There is a focusing effect, so that the particles deviating in the 
direction in which the field decreases tend to focus, while those going in the
direction of increasing field tend to de-focus.

ii) There are some particles with a given spin projection which deviate as 
those with a different spin projection. So, the Stern-Gerlach setup is 
not, even in theory, a ``completely reliable'' measuring apparatus. 

iii) There are some particles, with a definite spin projection along the
quantisation axis, which change the spin projection as they go through the 
magnet. So, the Stern-Gerlach setup is not an ``ideal'' measurement apparatus,
as successive measurements will not give exactly the same results.

This is what we will call the {\em semi-classical}
description of the Stern-Gerlach experiment.
Note that if we associate the particle position after the magnet as a 
``pointer'', which gives the result of the measurement of the spin projection
along the z-axis, then we conclude, that, in the {\em semi-classical} 
description, the Stern-Gerlach experiment is not an ideal measurement, 
because it can alter the spin projection, nor a completely reliable one, 
because the position is not always correlated with the spin projection.

These conclusions were obtained in a semi-classical framework, in which 
the motion of the particles was described by classical trajectories
which depended in the spin projection along the magnetic field that they 
encountered.  
Our motivation here is to see whether the same conclusions hold when the
full quantum mechanical problem is considered. In section 2
we formulate the time dependent quantum mechanical problem of a wave packet
going through a Stern-Gerlach magnet, and discuss the validity of the 
{\em text-book} and {\em semi-classical} approaches.
In section 3 and present the numerical solution of the quantum-mechanical 
problem. 
In section 4 we investigate several analytic 
approximations to the problem, considering the validity of the concept of
coherent internal states. In section 5 we discuss the interpretation of 
Stern-Gerlach experiments as measurements devices.
Section 6 is for summary and conclusions.

\section{Quantum mechanical formulation}

We want to investigate the effect of an inhomogeneous magnetic field on the
evolution of a quantum wave-packet. The situation that we will consider is a
magnetic field that has components in the $X$ and $Z$ directions, but not in
the $Y$ direction. This magnetic field has a length $L$, and it can be written
as
\begin{equation}
\vec B = (B_0 + B_1 Z) \vec u_z - B_1 X \vec u_x \quad, \quad 0\le Y \le L 
\quad.
\end{equation}
We use the capital letters $X,Y,Z,T$ to represent magnitudes with dimensions.
Low case $x,y,z,t$ correspond to dimensionless quantities.
We neglect border effects around $Y=0$ or $Y=L$. Note that this field fulfils
$\nabla \vec B =0$, and also $\nabla \times \vec B =0$, as it should be 
expected for a magnetic field in the region where there are no currents.
These conditions were not fulfilled in the case discussed in 
textbooks such as \cite{eis,lev,mer}.

The Hamiltonian which describes a non-relativistic neutral particle which 
enters in this field is given by
\begin{equation}
{\cal H} = {P_X^2 + P_Y^2 + P_Z^2 \over 2 M} - \mu \vec B \cdot \vec I \quad,
\end{equation}
where $\mu$ is the magnetic moment and $\vec I$ is the spin operator.

We consider now a wave packet $|\Psi(T);m_0\rangle$ which enters into this 
field. Initially, the wave packet can be characterised in coordinate space 
as a Gaussian 
 which is moving 
in the y-direction, while the initial spin projection along the $Z$-axis 
is $m_0$. 
\begin{equation}
\langle XYZ,m |\Psi(T=0);m_0\rangle = 
N \exp(- {X^2+Y^2+Z^2 \over 2 \sigma^2}) \exp(i k_y Y) \delta(m,m_0) \quad.
\end{equation}
Note that, neglecting the effects of the border, the $Y$-component of the 
wave-function is not affected by the interaction. On the other hand, the 
wave-function can be factorized into a $Y$-component and an $(X,Z)$-component. 
The
former component will evolve freely inside the magnet. Note that the 
wave-packet will stay within the magnetic field during a time 
$\tau=L / v_y$, where $v_y= \hbar k_y/M$. Assuming that the size of the wave-packet
$\sigma$ is very small compared to $L$, we can consider that the magnetic 
field starts at $T=0$, and finishes at $T=\tau$. So, we focus on solving the 
two-dimensional time dependent problem, which corresponds to calculate
the time evolution between the time $T=0$ and $T=\tau$ in a Hamiltonian
\begin{equation}
{ H} = {P_X^2  + P_Z^2 \over 2 M} - \mu \vec B \cdot \vec I,
\end{equation}
considering that the initial wave-function is 
\begin{equation}
\langle XZ;m|\Phi(T=0);m_0\rangle = N \exp(- {X^2+Z^2 \over 2 \sigma^2}) 
\delta(m,m_0) \quad.
\end{equation}
It is convenient to make use of dimensionless variables. So, we define
$x = X / \sigma$, $z = Z / \sigma$, $t = T/\tau$, $h  = H \tau / 
\hbar$. 
Then, the equation of motion becomes
\begin{equation}
h |\Phi(t);m_0\rangle = i {d \over d t}|\Phi(t);m_0\rangle .
\end{equation}
The dimensionless Hamiltonian can be written as $h = h_0 + v$, with
\begin{equation}
h_0 = {A \over 2}\left(p_x^2+p_z^2\right) \quad;\quad
v = - S \left(I_z (z + z_0) - I_x x \right) \quad.
\end{equation}
where $p_x = -i d/dx$, $p_z = -i d/dz$, and the dimensionless parameters $A, S, z_0$ are
\begin{equation} 
A={\hbar \tau \over  M \sigma^2} \quad;\quad
S ={\mu B_1  \tau \sigma \over \hbar } \quad;\quad 
z_0={ B_0   \over \sigma  B_1} \quad.
\label{defAS}
\end{equation}
The adiabaticity parameter $A$ is the ratio of the interaction time 
$\tau$ to the natural time of expansion
of the Gaussian packet.
The separation parameter $S$ is the ratio of the momentum 
change induced by the magnetic field gradient
divided by the momentum width of the Gaussian packet.
The  inhomogeneity parameter $z_0$ determines the relative change of the 
magnetic field in the range of the Gaussian.
Note that in the position $(x=0, z=-z_0)$, the magnetic field vanishes.
Note that the product $ AS = \mu B_1 \tau^2 / M \sigma $
is independent of $\hbar$. 
This magnitude is related 
to the deviation of the beam in the magnet.
For a given trajectory, which is determined by a fixed value of the product
$AS$, the {\em classical} limit is reached as $S\to \infty$ and $A \to 0$. 
Note that this corresponds to making $\hbar \to 0$ in eqs. \ref{defAS}.

\subsection{Validity of the semi-classical descriptions}

We will discuss the validity of the {\em semi-classical} and {\em text-book}
descriptions of the Stern-Gerlach experiment. It should be noticed that,
in general, a beam of particles is not given by a pure quantum mechanical
state, but rather by a mixture of small quantum wave packets. For definiteness,
we  consider that initially one has a distribution of particles
described as a Gaussian mixture, of range  $\sigma_m$,  of small Gaussian wave 
packets of range $\sigma$. The beam profile will then be characterised by
a Gaussian of range $\sigma_t=\sqrt{\sigma_m^2+\sigma^2}$.
The conditions required, in order to justify the {\em semi-classical} 
 description  done in \cite{prasara} are the following:

a) The inhomogeneity of the magnetic field over the quantum 
size of the wave packet should  be small: $\sigma  B_1 \ll B_0$. This implies 
that $z_0 \gg 1$.

b) The momentum change should be large compared to the quantum spread of the
beam momentum: $\mu B_1   \tau \gg {\hbar / \sigma}$. This implies that
$S \gg 1$. 

Note that these conditions are very well satisfied
in realistic situations for Stern-Gerlach experiments. 
However,
the validity of the {\em text-book} description requires also the far more 
stringent condition $ \sigma_t  B_1 \ll B_0 $, which require very strong
field $B_0$, or, alternatively, a very thin beam.

The purpose of this work is to investigate the full quantum solution of 
this problem for values of the parameters $z_0, S$ which are not necessarily
very large, so that the {\em semi-classical} and {\em text-book} description 
become dubious. Nevertheless, in order to have a reference to compare
the quantum calculation, we remind the expected results in the {\em text-book}
description.
The trajectory of the centre of the wave packet inside the magnet, 
is given by the expression
\begin{equation} 
z_m(t) = 1/2 (S A) m t^2 \quad.
\label{dev}
\end{equation}
which depends on the spin projection $m$. Note that, 
after the interaction ($t=1$), the positions of the
centre of the wave packets for each spin projection are given by 
$z_m(1)= SAm/2 $, and
their velocities are $\dot z_m(1) =  SAm/2 $. 
If, after the interaction, the beam
evolves freely during a time $t_d$, then the positions of the centre of the 
wave packets are expected to be given by
\begin{equation} 
z_m(t_d) = (1/2 + t_d) (S A) m  \quad.
\label{drift}
\end{equation}
As a typical value of the drift time $t_d$ we will consider the time necessary 
to reach the position $z_m=-z_0$, for the spin projection $m=-1/2$. Thus,
\begin{equation}
t_d = 2 z_0/(SA) - 1/2 \quad.
\label{deftd}
\end{equation}
Thus, we would expect that, after a drift time $t_d$, particles with spin
projection $m=1/2$ should appear around $z=z_0, x=0$, 
and particles with spin projection $m=-1/2$ should appear around $z=-z_0, x=0$.

\section{Numerical calculations}

We consider the scattering of a spin $1/2$ particle.
We expand the wave-function into two components, which have definite
spin projections along the z-axis.
\begin{eqnarray}
\langle xz; m=1/2 |\Phi(t);m_0\rangle &=& \alpha(x,z,t) e^{itS z_0/2} 
\nonumber \\ 
\langle xz; m=-1/2 |\Phi(t);m_0\rangle &=& \beta(x,z,t) e^{-itS z_0 /2} 
\end{eqnarray}
and the Schr\"odinger equation for the (x,z) plane can be written as
\begin{equation}\label{Sch}
\left[
\begin{array}{cc}
  {A \over 2}\left(p_x^2+p_z^2\right)-{S \over 2} z & {S \over 2} x \\
  {S \over 2} x & {A \over 2}\left(p_x^2+p_z^2\right) +{S \over 2} z \\
\end{array}
\right] \left[
\begin{array}{c}
 \alpha(x,z,t) \\
 \beta(x,z,t) \\
\end{array}
\right]=i {d \over dt} \left[
\begin{array}{c}
 {\alpha}(x,z,t) \\
 {\beta}(x,z,t) \\
\end{array}
\right]
\end{equation}
where $\alpha(x,z,t)$ and $\beta(x,z,t)$ are the components of the
spinor in the basis of the eigenstates of $I_{z}$. The numerical
solution of this equation has already been performed by Garraway
and Stenholm \cite{garraway}. However, they considered the case in which
$z_0$ was large, so their numerical result corresponded to the {\em text-book}
interpretation. A similar problem has been addressed by Franca et al 
\cite{franca}, but they made use of the adiabatic approximation, neglecting the
kinetic energy during the interaction time.

To follow our approach we must first write both components of
the spinor as linear combinations of harmonic oscillator
functions, so that
\begin{eqnarray}\label{desarrollo}
 \nonumber \alpha(x,z,t) &=&\sum_{nm}a_{nm}(t)\phi_{n}(x) \phi_{m}(z)\rangle
 \\
  \beta(x,z,t) &=&\sum_{nm}b_{nm}(t)\phi_{n}(x)\phi_{m}(z)
,
\end{eqnarray}
where $\phi_{n}(x)$ and $\phi_{m}(z)$ are the
harmonic oscillator eigenstates of order $n,m$ in the $x$ and in the
$z$ direction respectively. To
calculate the time-dependent coefficients $a_{nm}(t)$ and
$b_{nm}(t)$ of the expansion, it is natural to rewrite the
equation \ref{Sch} in terms of the well known creation and destruction
operators:
\begin{eqnarray}\label{Op}
  \nonumber a_{x}=\frac{1}{\sqrt{2}}(x+ip_{x}) \quad 
&& \quad   \nonumber a^\dagger_{x}=\frac{1}{\sqrt{2}}(x-ip_{x}) \\
   a_{z}=\frac{1}{\sqrt{2}}(z+ip_{z}) \quad 
&& \quad   a^\dagger_{z}=\frac{1}{\sqrt{2}}(z-ip_{z})
\end{eqnarray}
Thus, substituting the operators \ref{Op}  in eq. \ref{Sch}, 
we obtain the desired system of
ordinary coupled differential equations for the coefficients of the expansion
of $\alpha(x,z,t)$ and
$\beta(x,z,t)$:
 \begin{eqnarray}
\nonumber
\lefteqn{\dot{a}_{nm}=i\frac{A}{4}\left(a_{n+2,m}\sqrt{(n+1)(n+2)}+a_{n-2,m}\sqrt{n(n-1)}+\right.}\\
\nonumber && \left.a_{n,m+2}\sqrt{(m+1)(m+2)}+a_{n,m-2}\sqrt{m(m-1)}-2a_{nm}(n+m+1)\right)+ \\
\nonumber &&
i\frac{S}{2\sqrt{2}}\left(a_{n,m+1}\sqrt{m+1}+a_{n,m-1}\sqrt{m}-
\left(b_{n+1,m}\sqrt{n+1}+b_{n-1,m}\sqrt{n}\right)e^{-i S z_0 t }\right)\\
\nonumber && \rule{0pt}{15pt}\\
\nonumber \lefteqn{\dot{b}_{nm}=i\frac{A}{4}\left(b_{n+2,m}\sqrt{(n+1)(n+2)}+b_{n-2,m}\sqrt{n(n-1)}+\right.}\\
\nonumber && \left.b_{n,m+2}\sqrt{(m+1)(m+2)}+b_{n,m-2}\sqrt{m(m-1)}-2b_{nm}(n+m+1)\right)+ \\
  \nonumber && i\frac{S}{2\sqrt{2}}\left(-b_{n,m+1}\sqrt{m+1}-b_{n,m-1}\sqrt{m}-
\left(a_{n+1,m}\sqrt{n+1}+a_{n-1,m}\sqrt{n}\right)e^{i S z_0 t}\right),\\
\end{eqnarray}
where the dot stands for differentiation with respect to the
dimensionless parameter $t$. This system is solved
using a fourth order Runge-Kutta method. The number of harmonic oscillator 
basis functions needed in the calculation was typically of the order of 40
in each coordinate.

We have performed calculations using typical values of 
$A=0.5$, $S=4$, $z_0=4$. This corresponds to a case in which the
magnetic field vanishes at a distance of $4 \sigma$ . 
The time of the interaction is such that the width
of the beam would increase by a factor $\sqrt{1 + A^2}$. The magnetic field
gradient is such that each component of the magnetic field will acquire a 
momentum of $S \hbar / 2 \sigma$, in opposite directions.
As a comparison, we have also considered calculations with
$A=0.1$, $S=20$, $z_0=4$, that produce the same deviation of the
beam, but are closer to the classical limit. 

After the interaction, we consider a drift time $t_d$, given by eq.
(\ref{deftd}), during 
which the system evolves in the free Hamiltonian, so that the 
centre of the $m=\pm1/2$ wave-packet would  reach the point $z=\pm z_0$, 
according to the {\em text-book} description.

\begin{figure}[hbt]
\includegraphics[width=0.45\textwidth]{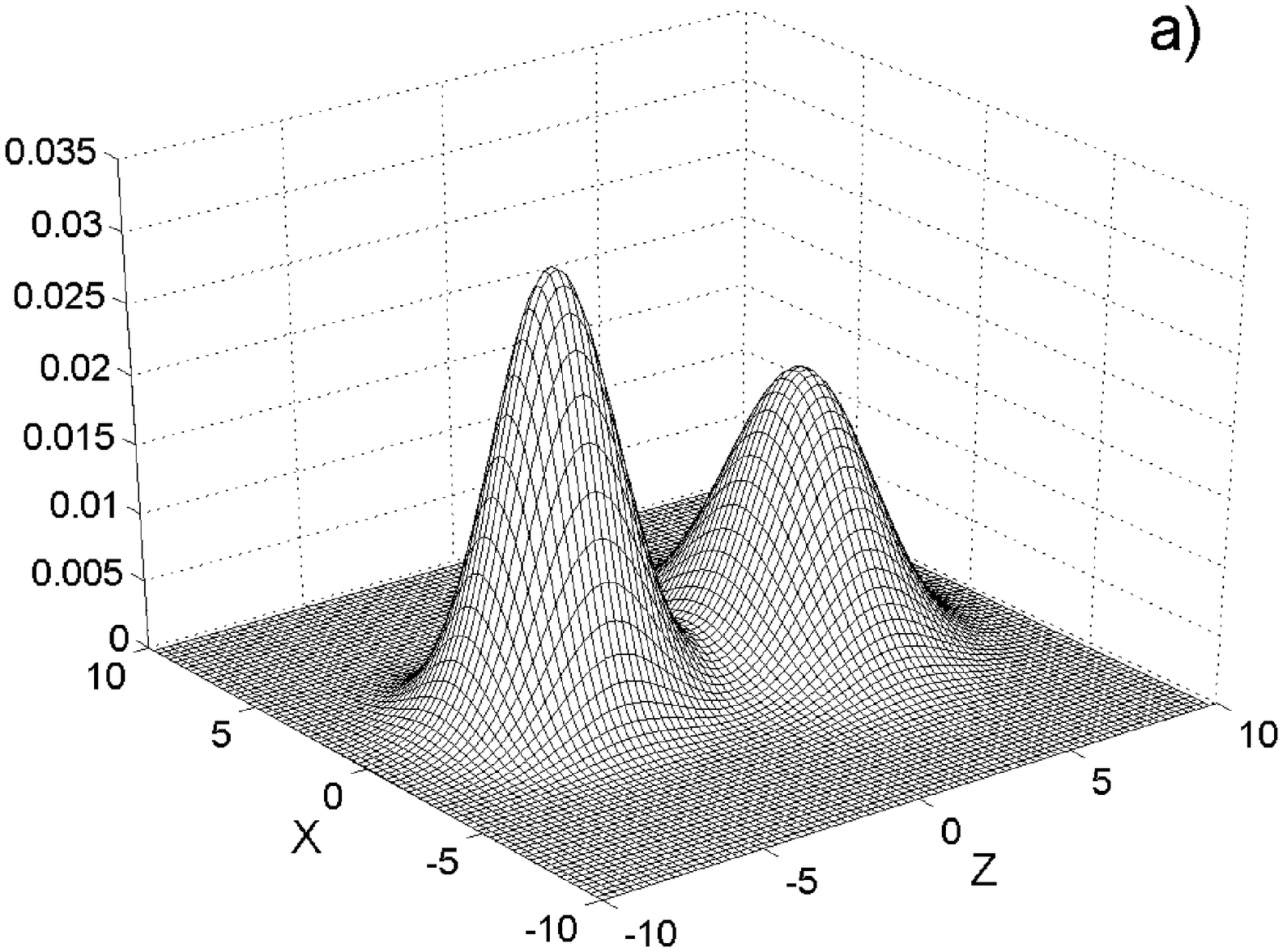}
\includegraphics[width=0.45\textwidth]{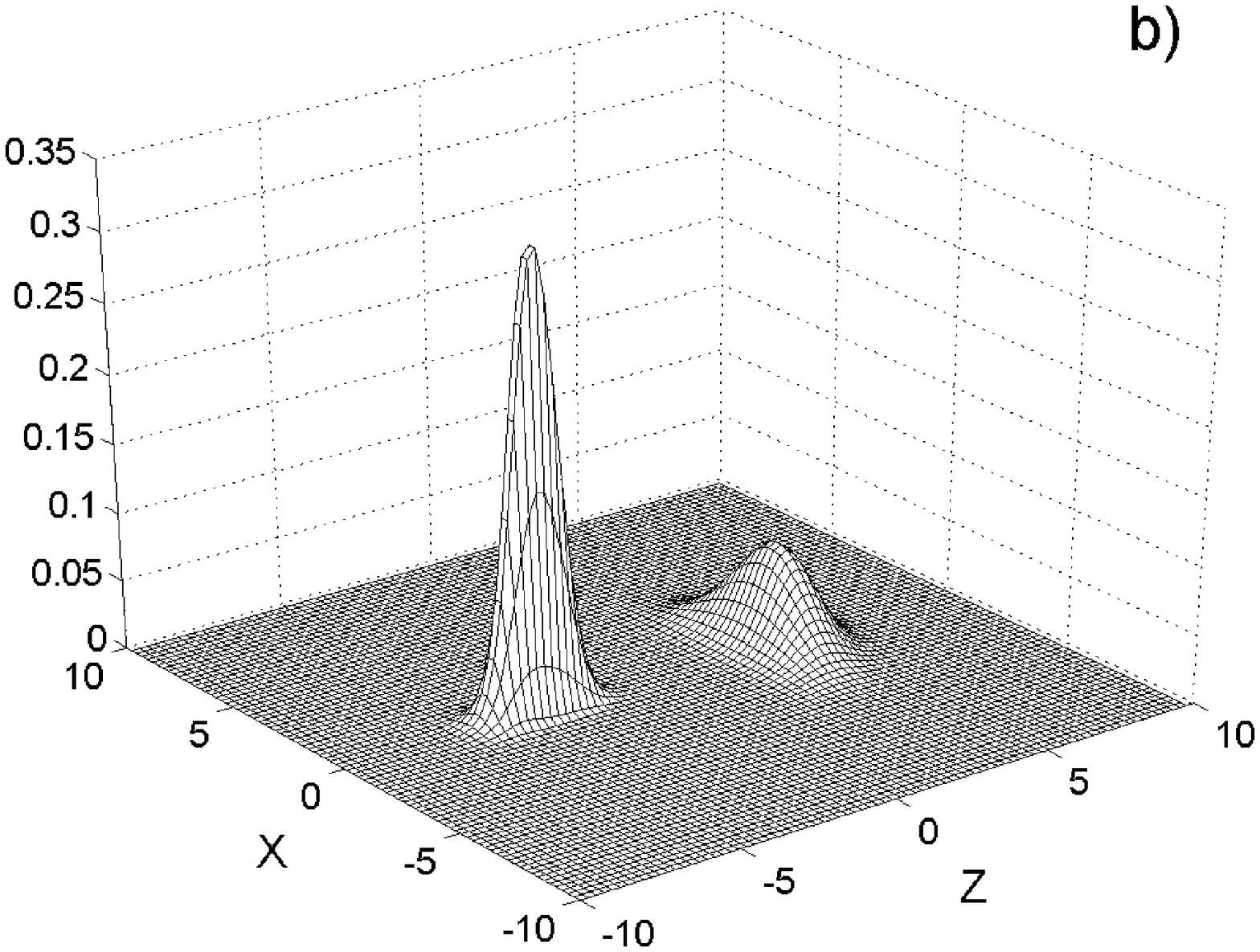}
\caption{Probability distribution for an unpolarised 
wave packet after going
through an inhomogeneous a magnetic field. Note the focusing effect of the 
lower component, which
corresponds predominantly to $m=-1/2$. The upper figure correspond to A=0.5,
S=4. The lower figure is for A=0.1, S=20, which is closer to the classical 
limit}
\label{focus}
\end{figure}

In figure \ref{focus} we represent the probability distribution of a wave 
packet, corresponding initially  to an unpolarised beam. This is given by
\begin{equation}
P_0(x,z) = {1 \over 2} \sum_{m m_0} |\langle x,z ; m |\Phi(t);m_0 \rangle|^2 
\end{equation}
The focusing effect 
can be clearly seen, by comparing the shape of the distributions for the
upper and lower components, which correspond predominantly to $m=1/2$ and
$m=-1/2$ respectively. The effect of the focusing is increased as $A$ decreases
and $S$ increases. 
So, we have confirmed that the focusing effect that was predicted in the 
semi-classical calculation in \cite{sara} is a genuine result, that appears 
in the quantum mechanical calculation, although it is diffused if the
adiabaticity parameter $A$ has a sizeable value.
It should be noticed that this focusing effect was also found in the 
calculations presented in \cite{garraway}.

\begin{figure}[hbt]
\includegraphics[width=0.5\textwidth]{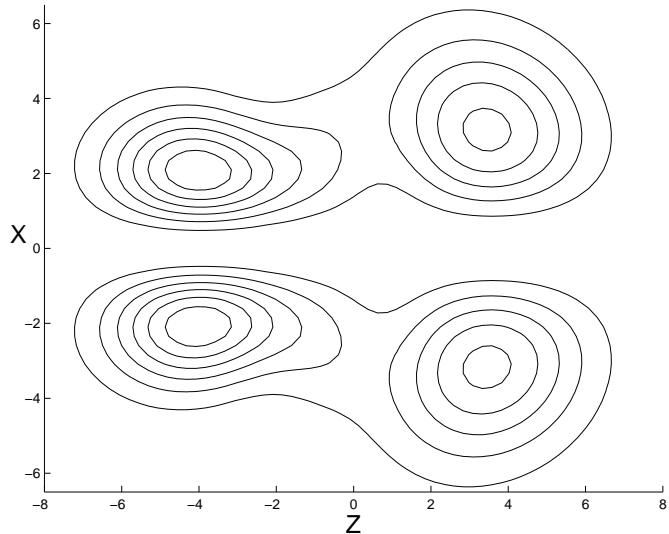}
\caption{Contour plot of the probability distribution of the spin flip 
component (spin up to spin down) 
of the wave-function. The maximum is $3.3\cdot 10^{-4}$}
\label{spinflip}
\end{figure}

In contrast to the {\em text book} description,
even if the initial beam has a definite spin projection 
along the z-axis, after the scattering process this spin projection can change.
We have evaluated the probability that the particles change their spin 
projection along the z-axis. It should be noticed that the probability of
going from spin up to spin down is not exactly the same of that of going from
spin down to spin up. For the reference case ($A=0.5, S=4, z_0=4$), we obtain 
that $p(1/2,-1/2)=0.0166$, $p(-1/2,1/2)=0.0198$. 

The spin-flip phenomenon also appears in the {\em semi-classical} description,
because not all the particles that compose the beam see the magnetic
field along the z-axis.
The {\em semi-classical} spin-flip probability is 
$p(1/2,-1/2)=p(-1/2,1/2)=0.0156$, which depends only
on the value of $z_0$. This is in good qualitative agreement with the
quantum calculations.
In Fig \ref{spinflip} we represent the spatial distribution of the spin
flip probability. Note that the spin flip probability  vanishes for
particles coming out along the z-axis.  
The spatial distribution of the spin-flip probability is in qualitative 
agreement with the semi-classical calculation, which becomes more accurate as
one makes the limit $A\to 0$, $S\to \infty$, with $AS$ constant.

The results of our calculations can be summarised as follows: When a beam of
particles, described by a Gaussian wave-function, and with a given 
spin projection along the z-axis goes through an inhomogeneous magnetic 
field, most of the 
particles scatter as expected in the {\em text-book} description. However,
a sizeable fraction of them, which depends on $z_0$ (about 2\% for $z_0=4$),
suffer a change of the spin projection (spin flip). From these particles that
suffer spin flip, about half scatter in the same direction as the majority of
the particles, and the other half scatter in the opposite direction.
We can conclude that the spin flip effect described in the 
{\em semi-classical}
description, which was not present in the {\em text book} description of
Stern-Gerlach experiments, is supported by the full quantum mechanical 
calculations. Also, we confirm that the Stern-Gerlach experiment, when 
considered as a measurement apparatus of the spin projection, is not an ideal
measurement (because there is spin-flip) and it is not fully reliable (because
there is not an exact correlation between the initial spin projection and 
the final position of the particle).

However, there are qualitative features of the full quantum
mechanical result, such as the difference
between up-down and down-up spin flip probabilities, that are not present in 
the {\em semi-classical} description and require further investigation.

\section{Approximate treatments}

Having solved numerically the problem, we will consider several approximate 
treatments to improve our understanding of the phenomena under consideration. 
The starting point is the exact 
evolution operator and the free evolution operator:
\begin{equation}
U(t)=\exp(-i (h_0 + v) t) \quad;\quad U_0(t)=\exp(-i h_0  t) \quad.
\end{equation}
It should be noticed that $h_0$ and $v$ do not commute.
We can use the following coordinates
\begin{equation}
\rho = \sqrt{(z+z_0)^2 + x^2} \quad,\quad \beta = \arctan{x \over (z+z_0)} \quad,
\end{equation}
and refer the spin components to the direction of the magnetic field at each
position
\begin{equation}
I_B = I_z \cos(\beta) - I_x \sin(\beta) \quad;\quad
I_T = I_z \sin(\beta) + I_x \cos(\beta) \quad. 
\end{equation}
In terms of these variables,  the initial state can be expressed as
\begin{equation}
\langle \rho \beta;m |\Phi(t=0);m_0\rangle = 
N \exp(-{\rho^2  -2 \rho z_0 \cos{\beta} +z_0^2\over 2}) \delta(m,m_0)
\end{equation}
and $h_0$ and $v$ take the expressions
\begin{equation}
h_0  =  {A \over 2} (p_\rho^2 + \rho^{-2} p_\beta^2), \quad
v  =  - S \rho I_B
\end{equation}
where $p_\rho$ and $p_\beta$ are the momenta associated to $\rho$ and $\beta$.
The relevant commutators are the following:
\begin{eqnarray}
\quad [h_0,v] &=& i AS (p_\rho I_B -  \{p_\beta,I_T\}/2 \rho)  \nonumber \\
\quad [[h_0,v],v] &=& - A S^2(I_B^2+I_T^2 - \{p_\beta, I_y\}/2 \\
\end{eqnarray}
Note that $[[h_0,v],h_0]=0$ and $[[[h_0,v],v],h_0]=0$.
For spin-1/2 particles, $I=1/2$,  $I_B^2=I_T^2=1/4$.

\subsection{Adiabatic Approximation}

The simplest approximation for the evolution operator consists in neglecting
completely the effect of $h_0$. This leads to the 
{\em adiabatic approximation},
given by
\begin{equation}
U(t) \simeq \exp(-it v) = \exp(it S \rho I_B)
\label{adiaU}
\end{equation}
Note that this expression conserves the projection of the spin along the 
direction of the magnetic field. Thus, it is convenient to expand the initial 
spin state into states $|n\rangle$ which fulfil $I_B |n\rangle = n |n\rangle$.
This can be done  considering the rotation of an angle $\beta$
around the y-axis  which takes the
$z-axis$ to the direction of the magnetic field. 
Thus, the adiabatic expression for the wave-function after the interaction
becomes
\begin{equation}
\langle \rho \beta ; m |\Phi(t);m_0\rangle = 
N   \exp(-{\rho^2  -2 \rho z_0 \cos{\beta} +z_0^2\over 2}) 
 \sum_n  d_{nm}^{1/2}(\beta) \exp(i n \rho St ) d_{nm_0}^{1/2}(\beta)
\label{adiawf}
\end{equation}
Note that this expression is equivalent to eq. (3.3) in \cite{franca}, where 
they expanded the wave-function in components that had definite spin 
projections along the local magnetic field. This expression contains the
qualitative features described in the numerical calculation. There is a 
spin-flip probability, as $m \ne m_0$. The focusing effect appears when this
adiabatic wave-function undergoes a free evolution during a time $t_d$ after
the interaction.
However, during the interaction time, the probability distribution is frozen.

\subsection{Pseudo-adiabatic Approximation}

The next approximation consists in neglecting the commutator $[h_0,v]$. This
leads to the {\em pseudo-adiabatic approximation}, given by
\begin{equation}
U(t) \simeq \exp(-it v) \exp(-i t h_0) = \exp(it S \rho I_B) U_0(t)
\label{pseudoadiaU}
\end{equation}
This expression also conserves the projection of the spin along the direction
of the magnetic field, but starting from a wave-function that has evolved
freely during the interaction time t. The wave-function has an analytic 
expression given by 
\begin{equation}
\langle \rho \beta ; m |\Phi(t);m_0\rangle = 
N  \exp(-{\rho^2 -2  \rho z_0 \cos{\beta} + z_0^2\over 2(1+iAt)}) 
 \sum_n  d_{nm}^{1/2}(\beta) \exp(i n \rho St) d_{nm_0}^{1/2}(\beta) 
\label{pseudoadiawf}
\end{equation}

The difference of this expression with the adiabatic one lies in the fact that
the Gaussian wave-packet gets wider during the interaction time, by a factor
$\sqrt{1+A^2}$, which is the widening of the free wave packet during the 
interaction time.

\subsection{Coherent State Approximation}

We consider the expansion of the evolution operator up to the third order
commutator. The following relations can be derived:
\begin{equation}
U(t) \simeq 
\exp( {(-i t)^3\over 6} [[h_0,v],v]])  
\exp(-itv)  \exp({(-it)^2\over 2} [h_0,v])    U_0(t)
\label{unosym}
\end{equation}
This expression is the basis for an analytic treatment of the wave-function.
For that purpose, we note that the dominant terms in the evolution operator are
those which conserve the spin projection along the direction of the magnetic 
field. The strongly oscillating factor $\exp(-itv)$ tends to cancel the terms
that do not conserve $I_B$.
Then, we retain in the expansion  only those terms which commute with
$I_B$. This leads to the expression:
\begin{equation}
U(t) \simeq \exp( - i t^3 A S^2 /12) 
\exp(it S \rho I_B) \exp(-i t^2 AS p_\rho I_B)  U_0(t)
\label{ucs}
\end{equation}
The operator $\exp(-i t^2 AS p_\rho I_B)$, when acting on
eigenstates of $I_B$, generates a 
displacement in  $\rho$, which is given by $\rho_f = \rho_i + t^2 AS I_B$.
This leads to an analytic expression for the wave-function, given by
\begin{equation}
\langle \rho \beta ; m |\Phi(t),m_0\rangle =  
\exp( i  A S^2 t^3/12)N 
\sum_n  d_{nm}^{1/2}(\beta)  \exp(i n \rho St) \sqrt{\rho_n \over \rho}
 \exp(-{\rho_n^2 -2  \rho_n z_0 \cos{\beta} - z_0^2\over 2(1+iAt)}) 
 d_{nm_0}^{1/2}(\beta)
\label{cswf}
\end{equation}
where $\rho_n=\rho-n ASt^2/2$. This wave-function conserves the spin projection  along the direction of the
magnetic field. Thus, the states with a definite spin projection along the
magnetic field in each position correspond to the coherent internal states
introduced in ref. \cite{prasara}. So, we call this approximation the 
{\em Coherent State approximation.} Note that in this approximation the
wave-function not only gets wider during the interacting region, but
 the components with different values of $I_B$ separate.

\subsection{Symmetrized approximation}

We can approximate the evolution operator by the following expression, which is
correct up to commutators of fourth order:
\begin{equation}
U(t)\simeq U_0(t/2) \exp(-itv - (-it)^3 [[h_0,v],v]/12) U_0(t/2)
\end{equation}

Neglecting the terms that do not commute with $I_B$, we have
\begin{equation}
U(t)\simeq\exp(  i t^3 A S^2 /24) U_0(t/2) \exp(it S \rho I_B) U_0(t/2)
\label{symU}
\end{equation}
The wave-function can be written as
\begin{equation}
|\Phi(t);m_0\rangle =\exp( i  A S^2t^3/24) U_0(t/2) |\Phi'(t);m_0 \rangle
\end{equation}
where
\begin{equation}
\langle \rho \beta ; n |\Phi'(t);m_0 \rangle =
N  \exp(-{\rho^2 -2  \rho z_0 \cos{\beta} + z_0^2\over 2(1+iAt/2)}) 
\sum_n d_{nm}^{1/2}(\beta) \exp(-i n \rho St)  d_{nm_0}^{1/2}(\beta)
\label{symwf}
\end{equation}
that, although it is not completely analytic, it can be applied to evaluate 
the expansion of the 
wave-function in a harmonic oscillator basis. 
This approximation corresponds to split the effect of $U_0(t)$ during the 
interaction symmetrically, taking half of it before and half of it after
the interaction.
Note that here also the evolution
associated to the interaction conserves the spin projection along the 
magnetic field. We call this the {\em symmetrized approximation}.

\subsection{Comparison with the exact calculation}

We have performed calculations with all the approximations. We find that the
qualitative characteristics of the exact calculations
discussed above, which are the focusing 
effect in the component which goes to negative z-values, and the presence of
spin-flip components, appear in all the calculations. The quantitative 
differences between the different approaches arise in the momentum 
distribution of the spin flip component. This comes out symmetric in the 
adiabatic and pseudo-adiabatic approximations (same probability distribution 
for positive and negative momenta), and not fully symmetric in the coherent 
state or symmetrized approximations, in closer agreement with the exact 
calculations.

\begin{figure}[hbt]
\includegraphics[width=0.5\textwidth]{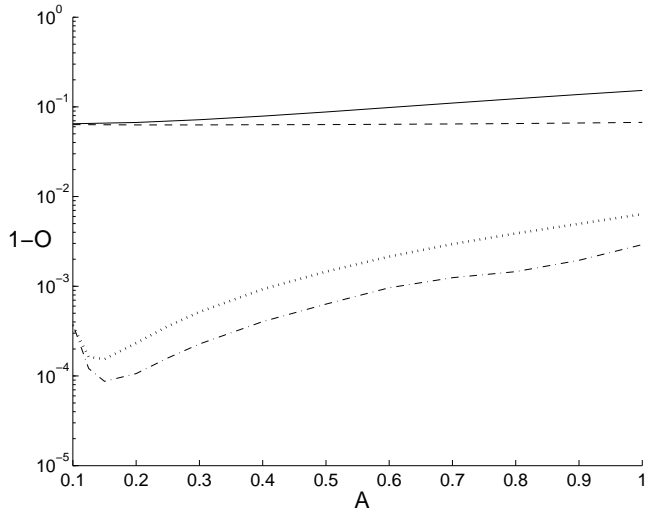}
\caption{Overlaps of the approximate wave-functions with the exact one, 
as a function of the adiabaticity parameter, for $SA=2$. The value $1-O=0$ 
correspond to perfect agreement. The full line is the adiabatic approximation,
the dashed line is the pseudo-adiabatic approximation, the dotted line is the
coherent state approximation and the dot-dashed line is the symmetrized approximation}  
\label{overlap}
\end{figure}

To evaluate the quality of these approximations, we have calculated the 
average of the overlap between the exact and the
approximate calculations. This overlap is defined as 
\begin{equation}
O = {1\over 2} \left| 
\sum_{m_0}  \langle 
\Phi_{\mbox{ex}}(t=1);m_0 |\Phi_{\mbox{ap}}(t=1);m_0 
\rangle \right|
\end{equation}
They are displayed in figure \ref{overlap}, as a function of the adiabaticity
parameter $A$, for a fixed value of the product $AS=2$, which determines the
deviation of the center of the wave packet in the magnetic field, as shown
in Eq. \ref{dev}. 
The quantity $1-O$ is about 10\%  for a wide range of values of $A$. In 
particular, for $A=0.5, S=4$, $1-0=0.088$ for the adiabatic calculation and
$1-O=0.064$ for the pseudo-adiabatic calculation.
On the contrary, the symmetrized and coherent state approximations are much 
better, so that $1-O$ is about 0.1\%. In 
particular, for $A=0.5, S=4$, $1-0=0.0015$ for the coherent state and
$1-0=0.0006$ for the symmetrized calculation. The reason for this better 
agreement arises from the fact that the coherent state and symmetrized 
calculation allow for the distortion in the wave-function produced by the 
magnetic field gradient, while for the adiabatic and pseudo-adiabatic 
calculation the effect of the field contributes only to a phase.

In all the calculations that we have performed, the quality of the
approximated calculations improves as one goes from the adiabatic, to the 
pseudo-adiabatic, to the coherent state and finally to the symmetrized
approximations. Globally considered, the approximations deteriorate as the 
product $SA$ gets larger, because then there is more distortion introduced
in the wave-function due to the combined effect of the interaction and the free
Hamiltonian.

A very interesting case is the limit $A\to 0$, $S\to \infty$,
for fixed values of $AS$.
Naively, one would expect that the adiabatic approximation would be adequate
here, as the free Hamiltonian $h_0$ is negligible compared to $v$.
However, this is not the case. As shown in figure
\ref{overlap}, the adiabatic and pseudo-adiabatic 
approximations are rather poor, giving values of $1-O$ about a few percent.
The coherent state and symmetrized approximations are very good for
$A = 0.015$, but then they become worse for smaller values of $A$. Numerical
calculations are very difficult when $S$ is large, because a large oscillator
basis is needed. An analytic solution of this limiting case would be desirable.

The interest of this limit case ($A\to 0$, $AS$ constant) is not only formal.
In nuclear physics there are cases in which weakly bound nuclei interact
strongly with targets during a very short time, so that the quantum state is 
significantly distorted. The validity of the adiabatic
approximation in these situations is open to debate \cite{ron}.

Note that in the definition of the overlap we allow for an overall phase 
difference between the exact and approximate wave-functions. This overall 
phase difference does not affect any observable. We find that the best 
approximate calculations (coherent state and symmetrized) only reproduce 
accurately the phase of the exact wave-function when both $A$ and $S$ are small.
This is apparently related to the effect of higher order terms in the 
commutator series of the evolution operator, which seem to affect only a global
phase in the wave-function.

So, we see from these approximations that a crucial feature of them is the fact
that the most relevant terms in the evolution operator conserve the spin 
projection along the local direction of the magnetic field. This is the basis 
of the semi-classical calculation performed in \cite{prasara}, in which the 
states with definite spin projections along the local magnetic field were 
taken as {\em coherent internal states}, and hence their motion could be 
described  in terms of trajectories.

Despite the fact that the approximations discussed here, specially the
coherent state and symmetrized approximations, are very accurate, they do not
describe an important effect of the exact evolution operator. In all the 
approaches described here, the scattering amplitudes for given spin projections
along the y-axis (the beam axis), are equal, up to a phase factors, to the 
amplitudes in which the spin projections are reversed. This is a result of the
fact that only terms which commute with $I_B$ are allowed in the expansion of
the evolution operator. 


\section{Revisiting Stern-Gerlach experiments}

In the {\em text book} description of the Stern-Gerlach experiment, the 
deflection 
of the beam gives information of the spin projection along the $z-$axis, which
is the one that points along the magnetic field at the centre of the beam.
The deflection of the beam is not sensitive to the spin components
along other directions. If, for a spin-1/2 particle, the initial spin points 
along the $x-$axis, $m_x=+1/2$, the 
{\em text book} description would indicate that the pattern of scattered 
particles would be completely equivalent to that one produced by a 
mixture of 50\% $m_z=+1/2$ and 50\% $m_z=-1/2$ 
particles. The same would be true for $m_x=-1/2$. So, a Stern-Gerlach 
experiment is not expected to give any asymmetry between different spin 
projections perpendicular to the z-axis.

\begin{figure}[hbt]
\includegraphics[width=0.45\textwidth]{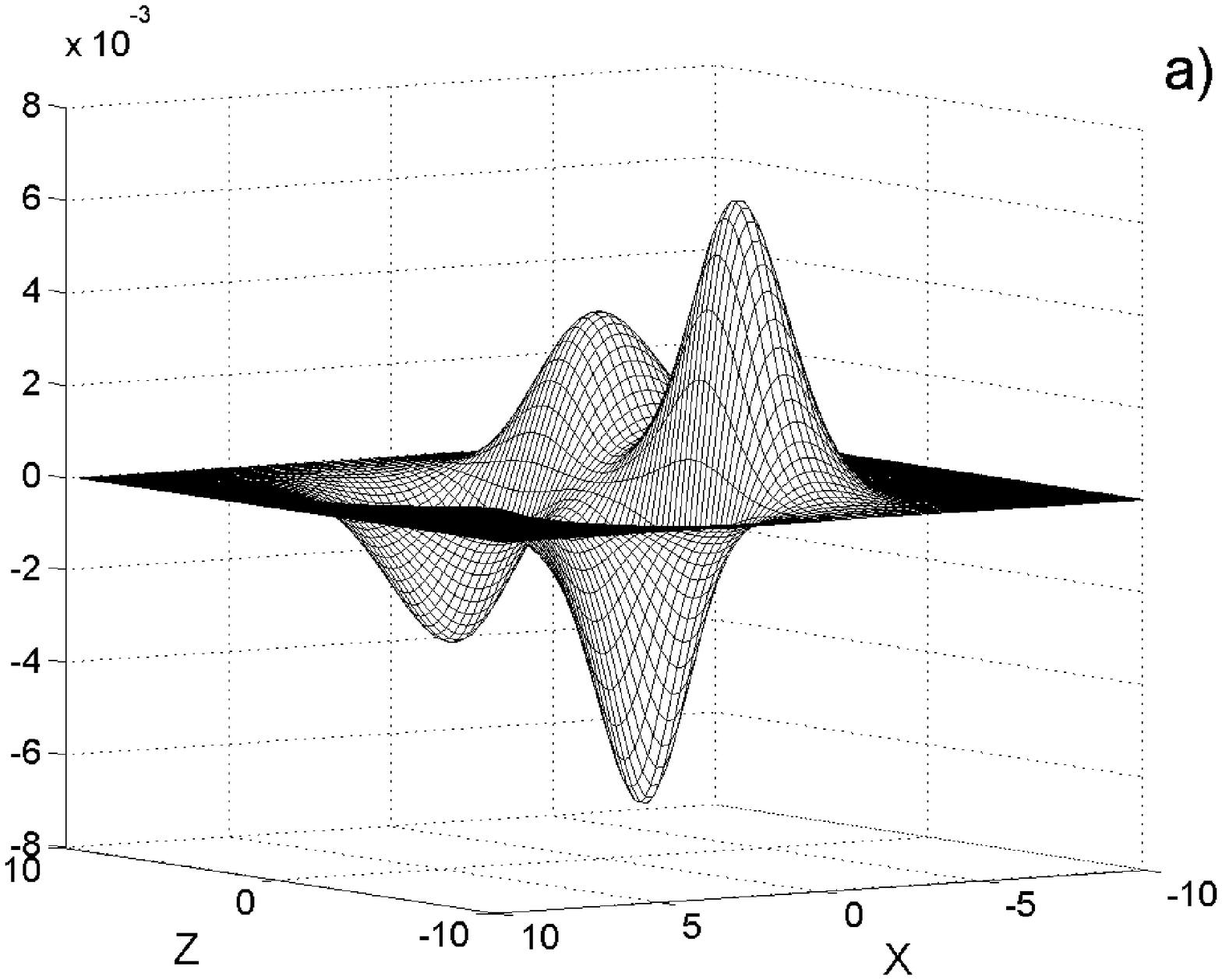}
\includegraphics[width=0.45\textwidth]{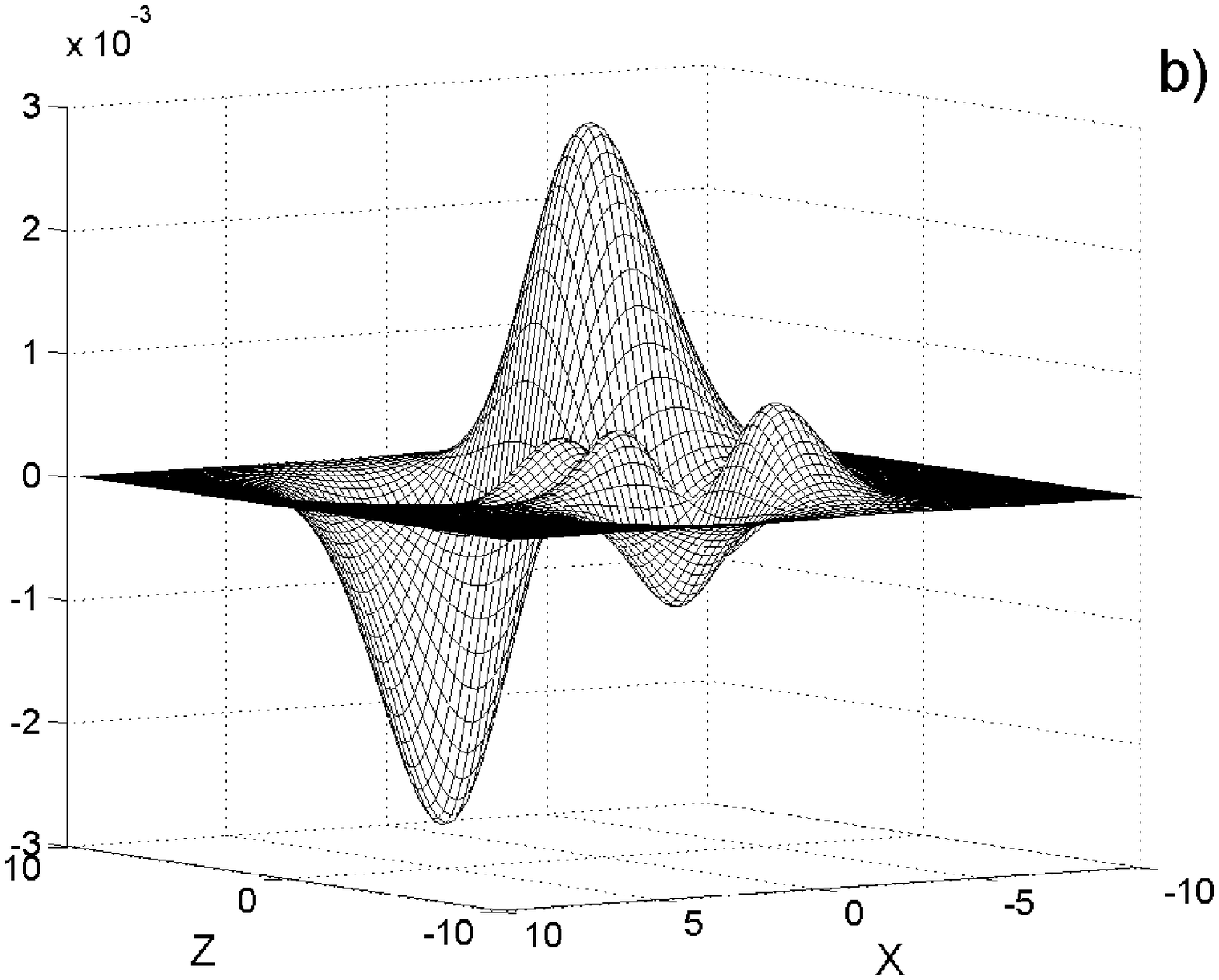}
\includegraphics[width=0.45\textwidth]{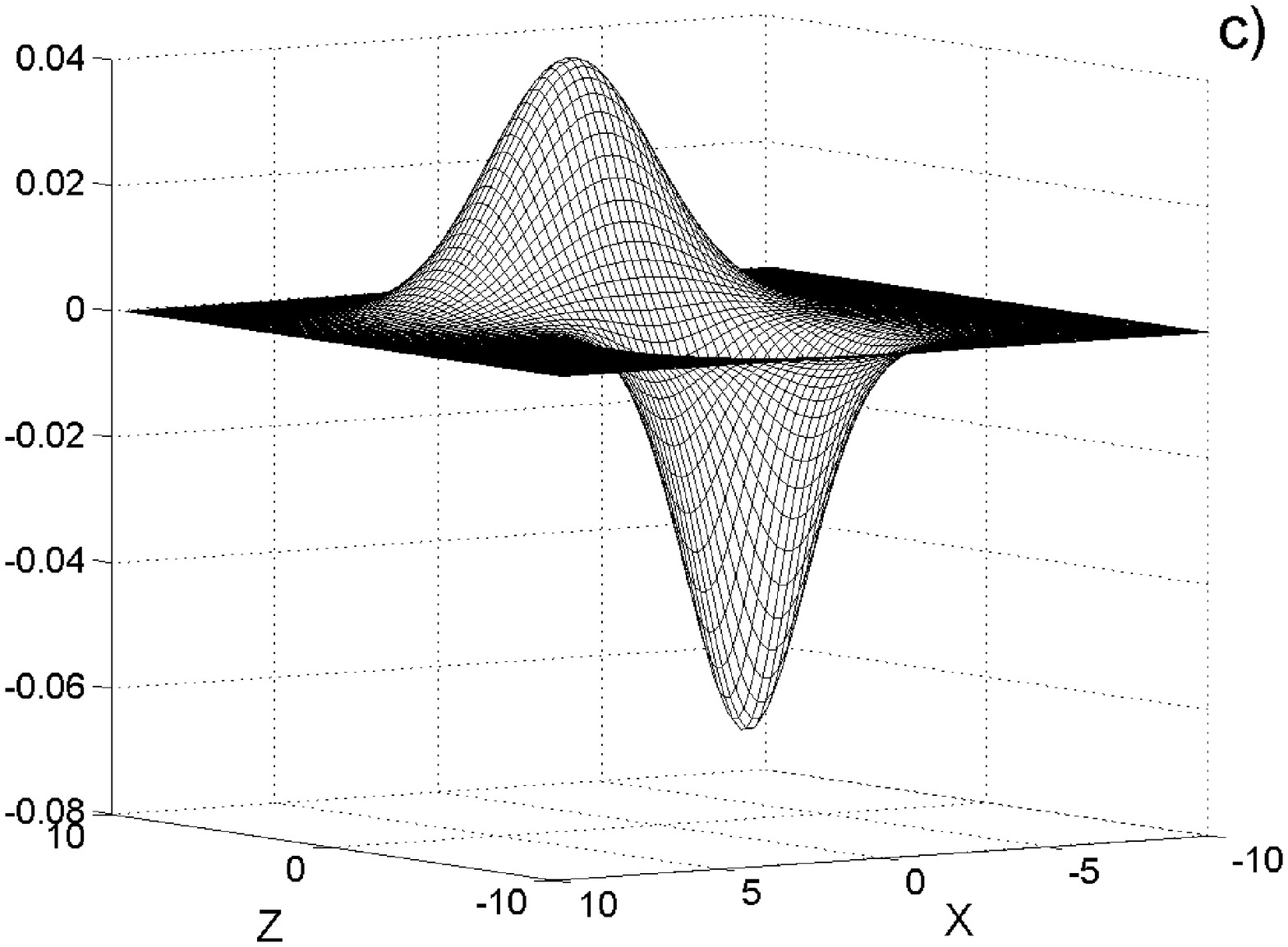}
\caption{Asymmetries for particles polarised  along the  
$y$ (a), $x$ (b), and $z$ (c) directions. 
Note that the maximum asymmetry occurs for
particles polarised along the z-axis, but that there are important
asymmetries for particles polarised along the x-axis and the y-axis. }
\label{asimetria}
\end{figure}

To investigate this question, we define the asymmetry for a given axis
as the difference in the probabilities of finding the scattered particles in
a given position in the $(z,x)$ plane, for the two spin projections. 
Thus, we have
\begin{eqnarray}
A_z(x,z) &=& \sum_{m m_0 m_0'} \langle x,z ; m |\Phi(t);m_0 \rangle 
 \langle x,z ; m |\Phi(t);m_0' \rangle^* \langle m_0|\sigma_z|m_0' \rangle \\
A_x(x,z) &=& \sum_{m m_0 m_0'} \langle x,z ; m |\Phi(t);m_0 \rangle 
 \langle x,z ; m |\Phi(t);m_0' \rangle^* \langle m_0|\sigma_x|m_0' \rangle \\
A_y(x,z) &=& \sum_{m m_0 m_0'} \langle x,z ; m |\Phi(t);m_0 \rangle 
 \langle x,z ; m |\Phi(t);m_0' \rangle^* \langle m_0|\sigma_y|m_0' \rangle 
\end{eqnarray}
Note that, in the standard description of the Stern Gerlach experiment,
the spin projection along the $z$ axis is conserved, and thus the asymmetries
$A_x$ and $A_y$ should vanish at all points.
This is not the case. As shown in figure \ref{asimetria}b, there is a 
difference in the pattern of particles scattered depending on the spin 
projection along the $x$-axis. This effect is found to depend 
on the inhomogeneity of the magnetic field, which is determined by
$z_0 = B_0 / B_1 \sigma$. If $z_0$ is large, the inhomogeneity of the 
magnetic field  explored by the beam is small, and so is $A_x$.
This asymmetry can be calculated, with various 
degrees of accuracy, making use of the approximate treatments discussed here.
It can also be calculated with the semi-classical treatment of \cite{prasara}.
The origin of this asymmetry can be understood by arguing that, the motion in
an inhomogeneous magnetic field conserves the spin projection along the 
local magnetic field, which has a different direction for the different parts
of the wave-function. This links with the concept of coherent internal states, 
which were introduced in \cite{sara}.

The calculations in figure \ref{asimetria}a show also that there is an asymmetry $A_y$ which means that 
there is a dependence of the spin projection along the y-axis. This is a
dynamical effect, which does not appear in the {\em semi-classical} description.
In fact, the analytic approximations presented here, the value of $A_y$ 
vanishes after the interaction. Only after allowing for some time of
free evolution, non-vanishing values of $A_y$ develop.
 The origin of this asymmetry arises from the term 
$A S^2 p_\beta I_y$ which appears in the double commutator $[[h_0,v],v]$.
The effect of this term can be understood because $p_\beta$ is the generator
of rotations in the $(x,z)$ plane, around the point $x=0, z=-z_0$ where
the field vanishes. The effect of this term in the expansion of the evolution 
operator, would generate a rotation in the wave-function, around the point
where the field vanishes, which will be opposite for the 
different spin projections along the z-axis. Indeed, this effect competes with
the interaction $v= S \rho I_B$, which tends to preserve the spin projection 
along the direction of the field. The result of this competition is that the
magnitude of the asymmetry depends on the ratio $AS/z_0$.

The fact that all the asymmetries are non-vanishing, and also that they have
different behaviour as a function of $(x,z)$, leads to an exciting 
possibility. Consider that we have a beam of particles, so that we do not
know their polarisation state. We can make the beam go through an 
inhomogeneous field, as described here, and detect the pattern of scattered 
particles. Let the polarisation state be described initially as a density 
matrix
$\rho = 1/2 (I + p_x \sigma_x + p_y \sigma_y + p_z \sigma_z)$, where $\vec p$
is a vector which measures the degree and direction of the beam polarisation.
Then, the density of particles detected in the $(x,z)$ plane will be 
proportional to 
\begin{equation}
P(x,z) = P_0(x,z)+ {1 \over 2}(p_x A_x(x,z) +p_y A_y(x,z)+p_z A_z(x,z))
\label{pxz}
\end{equation}
This allows to obtain all the components of the polarisation vector from the
pattern of scattered particles, when a sufficient number of particles are
detected.
Note that, in contrast to expression \ref{pxz}, the {\em text-book} 
description of The Stern-Gerlach experiment would be consistent with a
probability density given by
\begin{eqnarray}
P(x,z) &=& P_0(x,z)+ {1 \over 2} p_z A_z(x,z) \\
A_z(x,z) &=& 2 P_0(x,z) \quad z>0 \nonumber \\
A_z(x,z) &=& - 2 P_0(x,z) \quad z<0 \nonumber
\end{eqnarray}
This expression, when applicable, would allow to obtain  information 
only on the value of $p_z$. 

\section{Summary and conclusions}

We have investigated the motion of a particle with spin in an inhomogeneous 
magnetic field using a quantum mechanical framework. Our aim is to investigate 
in detail the limitations of the usual textbook approach to Stern-Gerlach
experiments, that assumes that the spin projection along the 
direction of the magnetic field is conserved, while different spin components 
acquire a momentum which depends on the gradient of the field.

We find that, consistently with a previous semi-classical analysis, there is
a sizeable probability of spin flip, which depends on the inhomogeneity of 
the field. Besides, there is a focusing effect in the component that deviates
towards in the direction in which the modulus of the field decreases. These
characteristics are very robust, and occur in dynamical situations which are 
far from the semi-classical limit.

Thus, we can conclude that the Stern-Gerlach experiment is not, even in 
principle, and ideal experiment, which would ``project'' the internal state
into the eigenvalues of the measurement operator. Moreover, the experiment is
not fully reliable, as the position or momenta of the particles do not give 
unequivocal information on the spin projection. The magnitude that determines
how close is a Stern-Gerlach experiment to an ideal reliable measurement is
$z_0 = \sigma B_0/B_1$. Only when the magnetic field $B_0$ is very large 
compared to its gradient, or when the size of the beam $\sigma$ is very small,
the Stern-Gerlach experiment would approximate to an ideal reliable measurement.

We have investigated different approximate treatments of the exact 
quantum-mechanical problem. We find that, to a good approximation, the 
interaction occurs as if the spin projection along the magnetic field at
each position was conserved. This indicates that, for each position in the
inhomogeneous field, the states with given spin projection along the
magnetic field are coherent internal states.
Then, provided that the quantum size of the wave-function is 
small compared to the inhomogeneity of the magnetic field, it is meaningful to 
approximate the motion of these states in terms of classical trajectories. 
This justifies the treatment performed in \cite{prasara}.

It is interesting to note that the adiabatic approximation it is not accurate,
even in the limit of small $A$ (large mass, or short interaction time), if, 
at the same time, the interaction is large so that it generates a fixed 
deflection angle. This observation can be relevant to cases, such as in 
nuclear physics \cite{ron},  in which although the collision times are short 
to guarantee
the validity of the adiabatic approximation, the forces are so strong to 
produce a finite deflection.

Our calculations indicate that the Stern-Gerlach experiment is not an ideal 
measuring apparatus, in the sense of \cite{neumann}. However, this does not 
mean   that one cannot acquire an accurate knowledge from the spin state of
the projectile by observing the statistical results of the experiment. On 
the contrary: while an idealised Stern-Gerlach experiment will not give any 
information of the spin projection along the $x$ or $y$ axis, the analysis 
of a realistic Stern-Gerlach experiment, such as modelled in our calculations,
can give the value of all the components of the density matrix that 
describes the polarisation  of the beam.

Our analysis supports the idea that the interpretation of realistic 
experiments does not require the use of the reduction principle,
as discussed by several authors in \cite{reduc}.
Thus, the interaction between the spin and the magnetic field, which is 
described in a purely quantum mechanical framework, generates a 
correlation between the spin polarisation of the beam and the final position 
of the particles of the beam. A measurement of a sufficiently large number
 of these positions, allows to determine the components of the density matrix
of the beam with sufficient statistical accuracy. The reduction principle 
is not required  in this argument.

\begin{acknowledgments}
 This work has been partially supported by the
Spanish MCyT, project FPA2002-04181-C04-04 
\end{acknowledgments}

\end{document}